\documentstyle[pra,aps,twocolumn]{revtex}
\begin{document}
\draft
\title{Congruences and Canonical  Forms for a Positive  
Matrix: Application to the Schweinler-Wigner Extremum Principle}
\author{R. Simon \thanks{email: simon@imsc.ernet.in}\\ The Institute of 
Mathematical Scinces, C. I. T. Campus, Chennai 600 113, India
\\
S. Chaturvedi \thanks{e-mail: scsp@uohyd.ernet.in} and V.
Srinivasan
\thanks {e-mail: vssp@uohyd.ernet.in}\\
School of Physics, University of Hyderabad, Hyderabad 500 046,
India}

\maketitle
\begin{abstract}
It is shown that a $N\times N$ real symmetric [complex
hermitian] positive definite matrix $V$ is congruent to a
diagonal matrix modulo a pseudo-orthogonal [pseudo-unitary]
matrix in $SO(m,n)$ [ $SU(m,n)$], for any choice of partition
$N=m+n$. It is further shown that the method of proof in this
context can easily be adapted to obtain a rather simple proof of
Williamson's theorem which states that if $N$ is even then $V$
is congruent also to a diagonal matrix modulo a symplectic
matrix in $Sp(N,{\cal R})$ [$Sp(N,{\cal C})$].  Applications of
these results considered include a generalization of the
Schweinler-Wigner method of `orthogonalization based on an
extremum principle' to construct pseudo-orthogonal and
symplectic bases from a given set of linearly independent
vectors.
 
\end{abstract}
\section{Introduction}
It is well known that a $N$-dimensional real symmetric [complex
hermitian] matrix $V$ is congruent to a diagonal matrix modulo
an orthogonal [unitary] matrix\cite{[1]}. That is, $V=S^\dagger
D S$ where $ D$ is diagonal and $S\in SO(N)$ [$S\in SU(N)$]. If,
in addition, $V$ is also positive definite, new possibilities
arise for establishing its congruence to a diagonal matrix.  For
$N$ even, it was shown by Williamson\cite{[2]} some sixty years
ago, and subsequently by several authors\cite{others,recent},
that such a $V$ is also congruent to a diagonal matrix  modulo a
symplectic matrix in $Sp(N,{\cal R})$ [$Sp(N,{\cal C})$].  That
is, $V>0$ implies $ V=S^\dagger D^\prime S$ where $ D^\prime$ is
diagonal and $S\in Sp(N,{\cal R})$ [$S\in Sp(N,{\cal C})$].
Williamson's theorem has recently been  exploited in defining
quadrature squeezing and symplectically covariant formulation of
the uncertainty principle for multimode states\cite{[3]}.  In
this work we establish yet another kind of congruence of a real
symmetric [complex hermitian] positive definite matrix to a
diagonal matrix valid, for both odd and even dimensions. We show
that an $N$-dimensional real symmetric [complex hermitian]
positive definite matrix $V$ is congruent to a diagonal matrix
modulo a pseudo-orthogonal [pseudo-unitary] matrix.  That is,
$V>0$ implies $V=S^\dagger D^{\prime\prime} S$ where $
D^{\prime\prime}$ is diagonal and $S\in SO(m,n)$ [$S\in
SU(m,n)$], for any choice of partition $N=m+n$.  A simple proof
of this result is given. The strategy adopted in proving this
result, with appropriate modification, works for the Williamson
case as well, and affords a particularly simple proof of
Williamson's theorem.  Needless to add that the diagonal entries
of neither $D^\prime$ nor $D^{\prime\prime}$ correspond to the
eigenvalues of $V$.
 
The theorems established here play a crucial role in enabling
one to construct pseudo-orthogonal and symplectic bases from a
given set of linearly independent vectors via an extremum
principle in the spirit of the work of Schweinler and
Wigner\cite{[5]}. In an important contribution to the age old
``orthogonalization problem'' -- the problem of constructing an
orthonormal set of vectors from a given set of linearly
independent vectors -- Schweinler and Wigner  proposed an
orthonormal basis which, unlike the familiar Gram-Schmidt basis
(which depends on the particular initial order in which the
given linearly independent vectors are listed), treats all the
linearly independent vectors on an equal footing and has since
found important application in wavelet analysis\cite{wavelet}.
More significantly, they showed that this special basis follows
from {\em an extremum principle}. In this work, we exploit our
results on congruence to obtain  generalizations of the
Schweinler-Wigner exremum principle leading to pseudo-orthogonal
and symplectic bases from a given set of linearly independent
vectors. Conversely, the extremum principle, once formulated,
can be interpreted as a procedure for finding the appropriate
congruence transformation to effect the desired diagonalization.

\section{Congruence of a positive matrix 
under pseudo-orthogonal [pseudo-unitary] transformations }

The fact that a real symmetric [complex hermitian] matrix is
congruent to a diagonal matrix modulo an orthogonal [unitary]
matrix  is  well known. While congruence coincides with
conjugation in the real orthogonal and complex unitary cases,
they become distinct when more general sets of transformations
are involved. A question which naturally arises is whether
congruence to a diagonal form can also be achieved through a
pseudo orthogonal [pseudo-unitary] transformation. The answer to
this question turns out to be in the affirmative with the caveat
that the matrix in question be  positive definite, and can be
formulated as the following theorem:

\noindent
{\it Theorem} 1: Let $V$ be a real symmetric positive definite
matrix of dimension $N$. Then, for any choice of partition
$N=m+n$, there exists an $S\in SO(m,n)$ such that
\begin{equation}
S^TVS=D^2={\rm diagonal}~({\rm and} >0).
\end{equation}
\vskip0.25cm
\noindent{\it Proof:}
We begin by recalling that the group $SO(m,n)$ consists of all
real matrices which satisfy $S^TgS=g,~ \mbox{det}\,S=1$, where
$g=$diag$(\:\underbrace{1,1,\cdots,1}_m\,,\,\underbrace{-1,\cdots,-1}_n\:)$.
Consider the matrix $V^{-1/2}gV^{-1/2}$ constructed from the
given matrix $V$. Since $V^{-1/2}gV^{-1/2}$ is real symmetric,
there exists a rotation matrix $R\in SO(N)$ which diagonalizes
$V^{-1/2}gV^{-1/2}$ :
\begin{equation}
R^TV^{-1/2}gV^{-1/2}R=\mbox{diagonal}\equiv \Lambda\,.
\end{equation}
This may be viewed  also as a congruence of $g$ using
$V^{-1/2}R$, and signatures are preserved under congruence.
(Indeed, signatures are the only invariants if we allow
congruence over the full linear group $GL(N,{\cal R})$ ). As a
consequence, the diagonal matrix $\Lambda$ can be expressed as
the product of a positive diagonal matrix and $g$ :
\begin{equation}
R^TV^{-1/2} g V^{-1/2} R = D^{-2} g = D^{-1} g D^{-1}\,.
\end{equation}
Here $D$ is diagonal and positive definite.

Taking the inverse of the matrices on both sides of (3) we find
that the diagonal entries of $gD^2 = D^2 g$ are the eigenvalues
of $V^{1/2} g V^{1/2}$ and that the columns of $R$ are the
eigenvectors of $V^{1/2} g V^{1/2}$.  Since $V^{1/2} g V^{1/2}$,
$gV$, and $Vg$ are conjugate to one another, we conclude that
$D^2$ is determined by  the eigenvalues of $gV \sim Vg$.

Define $S=V^{-1/2}RD$. It may be verified that $S$ satisfies the
following two equations :

\begin{eqnarray}
 S^TgS&=&g\,,\nonumber\\ S^TVS&=&D^2=\mbox{diagonal}\,.
\end{eqnarray}
The first equation says that $S\in SO(m,n)$ and the second says
that $V$ is diagonalized through congruence by $S$. Hence the
proof.

group $SO(m,n)$ by $SU(m,n)$, and $R\in SO(N)$ by $U\in SU(N)$
in the statement and proof of the above theorem, we have the
following theorem which applies to the  complex case.

\noindent
{\it Theorem} 2: Let $V$ be a hermitian  positive definite
matrix of dimension $N$. Then, for any partition $N=m+n$, there
exists an $S\in SU(m,n)$ such that
\begin{equation}
S^\dagger \,VS=D^2=~{\rm diagonal}~ ({\rm and}~>~0).
\end{equation}

\section{A simple proof of Williamson's theorem} 

It turns out that the above procedure when applied to the real
symplectic group of linear canonical transformations leads a
particularly simple proof of Williamsons's theorem.

\noindent 
{\it Theorem} 3: Let $V$ be a $2n$-dimensional real symmetric
positive definite matrix. Then there exists an $S\in Sp(2n,{\cal
R})$ such that
\begin{eqnarray}
S^TVS& =&D^2 >0\,,\nonumber\\
D^2&=&\mbox{diag}(\kappa_1,\kappa_2,\cdots,\kappa_n,\kappa_1,\kappa_2,\cdots,
\kappa_n).
\end{eqnarray}

\noindent
{\it Proof:} Note that the $2n$-dimensional diagonal matrix $D$
has only $n$ independent entries.  The group $Sp(2n,{\cal R})$
consits of all real matrices $S$ which obey the condition
\begin{equation}                          
S^T\beta S=\beta\,,~~~\beta= \left(\,
\begin{array}{cc}0 & 1 \\
-1 & 0\,\end{array}\,\right)\,,
\end{equation}
with $1$ and $0$ denoting the $n\times n$ unit  and zero
matrices respectively. Even though $S^T\beta S=\beta$ may appear
to suggest that $\mbox{det}S\, =\pm 1$, it turns out that $
\mbox{det}\,S = 1$. In other words, $Sp(2n,{\cal R})$ consists
of just one connected (though not simply connected) piece.
Indeed, for every $n\ge 1$ the connectivity property of $Sp(2n,
{\cal R})$ is the same as that of the circle.

The most general $S\in GL(2n,{\cal R})$ which solves $S^TVS=D^2$
is $S=V^{-1/2}RD$, where $R\in O(2n)$. Note that none of the
factors $D,R$ or $V^{-1/2}$ is an element of $Sp(2n,{\cal R})$.
However, a $V$-dependent choice of $D,R$ can be so made  that
the product $V^{-1/2}RD$ is an element of $Sp(2n,{\cal R})$ as
we shall now show.

Since $\beta^T=-\beta$, it follows that ${\cal M} =
V^{-1/2}\beta V^{-1/2}$ is antisymmetric. Hence there exists an
$R\in SO(2n)$ such that\cite{[6]}
\begin{equation}
R^T V^{-1/2} \beta V^{-1/2} R = \left(\,
\begin{array}{cc}
0 & \Omega\\ -\Omega & 0\,\end{array}\,\right),~~\Omega =
\mbox{diagonal} > 0\,.
\end{equation}
Define a diagonal positive definite matrix
\begin{equation}
D  =
\left(\,\begin{array}{cc}
\Omega^{-1/2} & 0\\
0 & \Omega^{-1/2}\end{array}\,\right)\,.
\end{equation}
Then we have
\begin{equation}
DR^TV^{-1/2} \beta V^{-1/2} RD = \beta\,.
\end{equation}
Now define $S=V^{-1/2}RD$. It may  be verified that  $S$ enjoys
the following properties: \begin{eqnarray} S^T\beta S &=&
\beta\,,\nonumber\\ S^TVS&=&D^2 ={\rm diagonal}.
\end{eqnarray}
The first equation says that $S\in Sp(2n, {\cal, R})$ and the
second one  says that $V$ is diagonalized by congruence through
S. This completes the proof of the Willianson theorem. To
appreciate the simplicity of the present the reader may like to
compare it with two recently published  proofs of the Williamson
theorem\cite{recent}.

We wish to explore the structure underlying the above proof a
little further so that the relationship between $D$ and $S$ in
(11) on the one hand and the eigenvalues and eigenvectors of
$\beta V^{-1}\,$(or $V^{-1/2}\beta V^{-1/2})$ on the other
becomes transparent.  Again consider the matrix ${\cal M} =
V^{-1/2}\beta V^{-1/2}$.  It is a real, non-singular,
anti-symmetric matrix and hence its eigenvalues $i\omega_\alpha$
and eigenvectors $\eta_\alpha$  have the following properties:
\begin{eqnarray}
{\cal M}\eta_\alpha &=& i\,\omega_\alpha
\eta_\alpha\,,~~~\alpha= 1,\cdots, 2n;\nonumber\\
\omega_k &>& 0\,,~~~k=1,\cdots,n\,;~~~~ \omega_{n+k} = -\omega_k\,;\nonumber\\
\eta_{n+k} &=& \eta_{k}^{*}\,;~~~~ k= 1,\cdots, n\,.
\end{eqnarray}

The eigenvectors $\eta_\alpha$ can be chosen to be  orthonormal
even when the eigenvalues $i\omega_\alpha$ are degenerate.
Arrange the eigenvectors $\eta_\alpha$ as columns of a matrix U.
The matrix $U$ thus obtained clearly belongs to the unitary
group $U(2n)$, and satisfies
\begin{equation}
U^\dagger{\cal M} U = \Lambda, ~~~
\Lambda=\left(\,\begin{array}{cc}i\Omega & 0\\
0 & -i\Omega\end{array}\,\right)\,,
\end{equation}
where $\Omega = {\rm diag}(\omega_1,\cdots,\omega_n) > 0$. Now
define the following $2n\times 2n$ unitary matrices
\begin{equation}
\Sigma= \left(\,\begin{array}{cc} 0 & 1\\
1 & 0 \end{array}\,\right),  ~~~~
\Delta = \frac{1}{{\sqrt 2}}\left(\,\begin{array}{cc} 1& -i\\
1& \,i\end{array}\,\right)\,.
\end{equation} 
These two matrices have the properties $\Sigma^2 = 1$,~
$U\Sigma= U^{*}$, and $\Sigma\Delta = \Delta^{*}\,$($^*$ denotes
complex cojugate of a matrix).  As a useful consequence of these
properties we have
\begin{equation}
U^{*}\Delta^{*} = U^{*} \Sigma \Sigma \Delta^{*} = U\Delta\,.
\end{equation}
We find that the unitary matrix  $U\Delta$ is real:  $U\Delta
\in O(2n)$.

Now consider $ S= V^{-1/2}U\Delta D$, where $D$ is a diagonal
matrix to be determined.  It follows from the definition of $S$
and the reality of $U\Delta\in O(2n)$ that
\begin{equation}
S^T V S = S^\dagger V S = D^2\,.
\end{equation}
Further, recalling that $U^\dagger{\cal M} U = \Lambda$ we
obtain
\begin{eqnarray}
S^T\beta S = S^\dagger\beta S &=& D\Delta^\dagger U^\dagger
{\cal M} U\Delta D
\nonumber\\
&=&D\Delta^\dagger \Lambda \Delta D = D
\left(\,\begin{array}{cc} O & \Omega\\ -\Omega & O
\end{array}\,\right)D\,.
\end{eqnarray}
It is now  evident that the following choice for $D$ ensures
that $S$ is an element of $S\in Sp(2n,{\cal R})$:
\begin{equation}
D= \left(\,\begin{array}{cc} \Omega^{-1/2} & O\\ O &
\Omega^{-1/2} \end{array}\,\right) \,.
\end{equation}
 
\noindent
This completes our analysis of the manner in which $S$ and $D$
are related to the eigenvalues and eigenvectors of the matrix
$\beta V^{-1}$.

 As in the pseudo-orthogonal case, by replacing the supercript
$^T$ by $^\dagger$ in the statement and proof of Theorem 3, one
obtains the following result.
 
\noindent 
{\it Theorem} 4: Let $V$ be a $2n$-dimensional hermitian
positive definite matrix. Then there exists an $S\in Sp(2n,{\cal
C})$ such that
\begin{eqnarray}
S^\dagger VS& =&D^2 >0\,,\nonumber\\
D^2&=&\mbox{diag}(\kappa_1,\kappa_2,\cdots,\kappa_n,\kappa_1,\kappa_2,\cdots,
\kappa_n).
\end{eqnarray}

An immediate consequence of the theorems stated above is that
for a real symmetric [complex hermitian] positive definite
matrix we can not talk about {\em the} canonical form under
congruence, for there are $m+n$ possible choices of $SO(m,n)$
[$SU(m,n)$], and in the case of even dimension one more choice
coming from Williamson's theorem. Needless to add that for the
same matrix $V$, the diagonal matrix  $D$  will be different for
different choices.

\section{Orthogonalzation Procedures}
Assume that we are given a set of linearly independent
$N$-dimensional vectors $v_1,\cdots,v_N$. Let $G$ denote the
associated Gram matrix of pairwise inner products: $ G_{ij}=
(v_i,v_j)$. The Gram matrix is hermitian by construction, and
positive definite by virtue of the linear independence of the
given vectors. The orthogonalization problem, i.e., constructing
a set of orthonormal vectors out of the given set of linearly
independent vectors, amounts to finding a  matrix $S$  that
solves
\begin{equation} 
S^\dagger G S = 1,~~\bbox{i.e.},~ G^{-1} = SS^{\dagger}\,.
\end{equation}
Each such $S$ defines an orthogonalization procedure.

Let us arrange the set of $N$ vectors as the entries  of a row
${\bf v} = (v_1,v_2,\cdots,v_N)$, and let ${\bf z} =
(z_1,z_2,\cdots,z_N)$ represent a generic orthonormal basis. The
orthonormal set of vectors {\bf z} corresponding to a chosen $S$
are related to the given set of linearly independent  vectors
through ${\bf z}= {\bf v}S $.  Clearly, there are infinitely
many choices for $S$ satisfying $(20)$: given an $S$ satisfying
$(20)$, any $S^\prime = SU$ where $U$ is an arbitrary unitary
matrix also satifies $(20)$. Thus the freedom available for the
solution of the orthonormalization problem is exactly as large
as the unitary group $U(N)$, and this was to be expected.

Schweinler and Wigner\cite{[5]} posed and answered the following
question: is there a way of descriminating between various
choices of $S$ that solves (20) and hence between various
orthogonalization procedures? They argued that a particular
choice of orthogonalization procedure should correspond
ultimately to the extremization of a suitable scalar function
over the manifold of all orthonormal bases, with the given
linearly independent vectors appearing as parameters in the
function. Different choices of onthonormal bases will then
correspond to different functions to be extremized. They
preferred the function to be symmetric under permutation of the
given vectors.  As an example they considered the following
function which is quartic in the given vectors:
\begin{equation}
gm({\bf z})=\sum_{k}\left(\sum_{l} {\mid (z_k,v_l)\mid }^2
\right)^2\,.
\end{equation}
They  showed that the extremum (maximum in this case) value of
$m({\bf z})$ is given by ${\rm tr}(G^{2})$, and this value
corresponds to the orthonormal basis ${\bf z} ={\bf v} U_0
P^{-1/2}$, where $U_0$ is the unitary matrix which diagonalizes
$G$: $U_0\:^\dagger G U_0 = P $.  We may refer to this  as the
Schweinler-Wigner basis, and the function $m({\bf z})$ as the
Schweinler-Wigner quartic form. It is clear that $U_0$ and hence
the Schweinler- Wigner basis is essentially unique if the
eigenvalues of the Gram matrix $G$ are all distinct.  We may
note in passing that, unlike the Gram-Schmidt orthogonalization
procedure, the Schweinler-Wigner procedure is democratic in that
it treats all the linearly independent vectors ${\bf v}$ on an
equal footing.

The content of the work of Schweinler and Wigner has recently
been reformulated\cite{[7]} in a manner that offers a clearer
and more general picture of the Schweinler-Wigner quartic form
$m{(\bf z)}$ and of the orthonormal basis which maximizes it.
This perspective on the orthogonalization problem plays an
important role in our generalizations of the Schweinler-Wigner
extremum principle, and hence we summarise it   briefly.

Since every orthonormal basis is the eigenbasis of a suitable
hermitian operator, it is of interest to characterize the
Schweinler-Wigner basis in terms of such an operator. Given
linearly independent $N$-dimensional vectors ${\bf v} =
(v_1,v_2,\cdots,v_N)$, the operator
$\hat{M}=\displaystyle{\sum_j} v_jv_j^\dagger$ is hermitian
positive definite. In a {\it generic orthonormal} basis ${\bf
z}$, it is represented by a hermitian positive definite  matrix
$M({\bf z}):\; M({\bf z})_{ij} = (z_i, \hat{M}z_j)$. Under a
change of orthonarmal basis ${\bf z}\rightarrow {\bf z}^\prime =
{\bf z}S$, $M({\bf z})$ transforms as follows
\begin{equation}
M({\bf z}) \to M({\bf z}') = S^\dagger M({\bf z})S\,,\,\,\; S
\in U(N)\,. \end{equation} Recall that $U(N)$ acts transitively
on the set of all orthonormal bases   and that ${\rm tr}(M({\bf
z})^2)=\displaystyle{\sum_{j,k}}|M({\bf z})_{jk}|^2$ is
invariant under such a change of basis, and hence is endependent
of ${\bf z}$. The Schweinler-Wigner quartic form $m({\bf z})$
can easily be identified as $\displaystyle{\sum_k} (M({\bf
z})_{kk})^2$. In view of the above invariance, maximization of
$\displaystyle{\sum_k} (M({\bf z})_{kk})^2$ is the same as
minimization of $\displaystyle{\sum_{j\ne k}} |M({\bf
z})_{jk}|^2$. The absolute minimum of $\displaystyle{\sum_{j\ne
k}}|M({\bf z})_{jk}|^2$ equals zero, and obtains when $M({\bf
z})$ is diagonal. Thus, the orthonormal basis which maximizes
$\displaystyle{\sum_k} (M({\bf z})_{kk})^2$ is the same as the
one in which $\hat{M}$ is diagonal, and we arrive at the
following important conclusion of Ref.[9]:

\noindent
{\em Theorem} 5: The distinquished orthonormal basis which
extremizes the Schweinler-Wigner quartic form $m({\bf z})$ over
the manifold of all orthonormal bases is the same as the
orthonormal basis in which the positive definite matrix $M({\bf
z})$ becomes diagonal.

Important for the above structure is the fact that the invariant
${\rm tr}(M({\bf z})^2)$ is the sum of  non-negative quantities,
and therefore a part of it is necessarily bounded. It is
precisely this property, which can be traced to the underlying
unitary symmetry, that is not available when we try to
generalize the Schweinler-Wigner procedure to construct
pseudo-orthonormal and symplectic bases wherein the underlying
symmetries are the noncompact groups $SO(m,n)$ and $Sp(2n,{\cal
R})$ respectively..

\section{ Lorentz basis with an extremum property}  
In this Section we show how the Schweinler-Wigner procedure  can
be generalized to construct pseudo-orthonormal basis based on an
extremum principle. We begin with the case of real vectors.

We are given a set of linearly independent real $N$-dimensional
vectors  ${\bf v} = (v_1,\cdots,v_N)$ and we want to construct
out of it a pseudo-orthonormal basis [$SO(m,n)$  Lorentz  basis
with $N=m+n$], i.e., a set of   vectors ${\bf z}=(z_1,
\cdots,z_N)$ satisfying
\begin{equation}   
(z_k, gz_l) = g_{kl}\,,~~g=
\mbox{diag}(\:\underbrace{1,1,\cdots,1}_m\,,\,\underbrace{-1,\cdots,-1}_n\:).
\end{equation}
Let $\hat{M}=\displaystyle{\sum_j} v_j v_j ^T$ as before, and
let the symmetric positive definite matrix $M({\bf z}):~~M({\bf
z})_{ij} = (z_i, \hat{M}z_j)$ represent $\hat{M}$ in a {\em
generic pseudo-orthonormal} basis ${\bf z}$. Under a
pseudo-orthogonal change of basis ${\bf z} \to {\bf z}' = {\bf
z}S$, the matrix $M({\bf z})$ transforms as follows:
\begin{equation}
M({\bf z}) \to M({\bf z}') = S^T M({\bf z})S\,,~~ S\in
SO(m,n)\,.  \end{equation} Since $S^T g S = g$ (or $gS^T  =
S^{-1}g$) by definition, we have
\begin{equation}
S:~~gM({\bf z}) \to gM({\bf z}') = S^{-1}gM({\bf z})S.
\end{equation}
That is, as $M({\bf z})$ undergoes congruence, $gM({\bf z})$
undergoes conjugation. Thus, ${\rm tr}(gM({\bf z}))^l$,
$l=1,2,\cdots,$ are invariant. In what follows we shall often
leave implicit the dependence of $M$ on the generic
pseudo-orthonormal basis ${\bf z}$.

Consider the invariant ${\rm tr}(gM({\bf z})gM({\bf z}))$
corresponding to $l=2$.  Write $M=M^{\rm even} + M^{\rm odd}$
where
\begin{equation}
M^{\rm even} = {1\over2}(M+gMg)\,,\,\,M^{\rm odd} =
{1\over2}(M-gMg)\,.
\end{equation}
In the above decomposition we have exploited the fact that $g$
is,  like parity,  an {\em involution}.

With $M$ expressed   in the $(m,n)$ block form
\begin{equation}
M=\left(\,\begin{array}{cc} A & C\\ C^T &
B\end{array}\,\right)\,,\,\; A^T=A\,,\,\,B^T=B\,,
\end{equation}
we have
\begin{equation}
M^{\rm even}= \left(\,\begin{array}{cc} A & 0\\ 0 &
B\,\end{array}\right)\,,\,\; M^{\rm odd} =
\left(\begin{array}{cc} 0 & C\\ C^T & 0\,\end{array}\right)\,.
\end{equation}

Symmetry of $M$ implies that  $M^{\rm odd}$ and $M^{\rm even}$
are symmetric.  Further, $M^{\rm odd}$ and $M^{\rm even}$ are
trace orthogonal: ${\rm tr}(M^{\rm odd} M^{\rm even})=0$. Thus,
\begin{equation}
{\rm tr}(gMgM) = {\rm tr}(M^{\rm even})^2 - {\rm tr}(M^{\rm
odd})^2\,,
\end{equation}
which can also be written as
\begin{equation}
{\rm tr}(MgMg) = {\rm tr}(M^2) - 2 {\rm tr}(M^{\rm odd})^2\,.
\end{equation}
A few observations are in order:
\begin{itemize}
 
\item In contradistinction to the original unitary case, the invariant in the 
present case is no more a sum of squares.  This can be traced to
the non-compactness of the underlying $SO(m,n)$ symmetry. As one
consequence, $\displaystyle{\sum_k} (M_{kk})^2$ is not bounded.
As an example, consider the simplest case $m=1,\;n=1$ and let
\begin{equation}
M= \left(\,\begin{array}{cc}a & 0\\0 &
b\,\end{array}\right)\,,~~a, b > 0.
\end{equation} 
Under congruence by the $SO(1,1)$ element
\begin{equation}
S= \left(\,\begin{array}{cc}
\cosh\mu & \sinh\mu\\
\sinh\mu & \cosh\mu\,\end{array}\right)\,,
\end{equation}
the value of $\displaystyle{\sum_k} (M_{kk})^2$ changes from
$a^2 + b^2$ to $ a^2 +b^2 + 2ab\sinh^2\mu \cosh^2 \mu$, which
grows with $\mu$ without bounds, showing that
$\displaystyle{\sum_k} (M_{kk})^2$ and hence $\mbox{tr} (M^2)$
is not bounded. Thus, in contrast to the unitary case,
extremization of the Schweinler-Wigner quartic form
$\displaystyle{\sum_k} (M_{kk})^2$ will make no sense in the
absence of further restrictions.

\item  The structure of the invariant ${\rm tr}(gMgM)$ in (30)  suggests 
the further restriction needed to be imposed:  within the
submanifold of pseudo-orthogonal bases ${\bf z}$ which keep
${\rm tr}(M({\bf z})^{\rm odd})^2$ (and hence ${\rm tr}(M({\bf
z})^2)$) at a fixed value we can maximize
$\displaystyle{\sum_k}M({\bf z})^2_{kk}$. In particular we can
do this within the submanifold which minimizes ${\rm tr}(M({\bf
z})^{\rm odd})^2$, and hence ${\rm tr}(M({\bf z})^2)$.  Clearly,
zero is the absolute minimum of the nonnegative object ${\rm
tr}(M({\bf z})^{\rm odd})^2$.  But by theorem 1 there exists a
Lorentz basis ${\bf z}$ in which $M({\bf z})$ is diagonal and
hence $M({\bf z})^{\mbox{odd}} = 0$. Thus the minimum ${\rm
tr}(M({\bf z})^{\rm odd})^2=0$, and hence the minimum of ${\rm
tr}(M({\bf z})^2)$, namely ${\rm tr}(gM({\bf z})gM({\bf z}))$,
is attainable.
\end{itemize}

The above observations suggest the following {\em two step
analogue of the Schweinler-Wigner extremum principle for Lorentz
bases}.  Choose the submanifold of Lorentz bases which minimize
the quartic form ${\rm tr}(M({\bf z})^{\rm odd})^2$, and
maximize the Schweinler-Wigner quartic form $ m({\bf z}) =
\displaystyle{\sum_k} (M({\bf z})_{kk})^2$ within this
submanifold.  Clearly, the first step takes $M$ to a
block-diagonal form, and the second one diagonalizes it.  Thus
we have established the following generalization of Theorem 5 to
the pseudo-orthonormal case:

\noindent
{\em Theorem} 6: The distinquished pseudo-orthonormal basis
which extremizes the ``Schweinler-Wigner'' quartic form $m({\bf
z})$ over the submanifold of pseudo-orthonormal bases which
minimize the quartic form $\mbox{tr}(M({\bf z})^2)$ is the same
as the pseudo-orthonormal basis in which the positive definite
matrix $M({\bf z})$ becomes diagonal.

 The submanifold under reference consists of Lorentz bases which
are related to one another through the maximal compact
(connected) subgroup of $SO(m,n)$, namely $SO(m)\times SO(n)$.
This subgroup consists of matrices of the block-diagonal form
\begin{equation}
\left(\begin{array}{cc}
R_1 & 0\\ 0 & R_2\end{array}\right)\,,\;\;\; R_1 \in
SO(m)\,\,,\;\;R_2\in SO(n)\,,
\end{equation}
and this is precisely the subgroup of $SO(m,n)$ transformations
that do not mix the even and odd parts of $M({\bf z})$.

To conclude this Section we may note that the above construction
carries over to the complex case, with obvious changes like
replacing $^T$ by $^\dagger$ and $SO(m,n)$ by $SU(m,n)$.

 \section{\bf Sympletic Basis with an Extremum Property}

Our construction in the pseudo-orthogonal case suggests a scheme
by which the Schweinler-Wigner extremum principle principle can
be generalized to construct a symplectic basis.  Suppose that we
are given a set of linearly independent vectors $ {\bf
v}=(v_1,v_2,\cdots,v_{2n})$ in ${\cal R}^{2n}$. The natural
symplectic structure in $R^{2n}$ is specified by the standard
symplectic ``metric'' $\beta$ defined in (7).  Let ${\bf z} =
(z_1, z_2, \cdots, z_{2n})$ denote a generic symplectic basis.
That is, $(z_j,\beta z_k)=\beta _{jk}\,,\,\,j,k=1,2,\cdots,2n$.
The real symlectic group $Sp(2n,R)$ acts transitively on the set
of all symplectic bases.

To generalize the Schweinler-Wigner principle to the symplectic
case, we begin be defining
$\hat{M}=\displaystyle{\sum_{j=1}^{2n}} v_j v_j^T$.  Let $M({\bf
z}):~M({\bf z})_{ij} = (z_i, \hat{M}z_j)$ be the symmetric
positive definite matrix representing the operator $\hat{M}$ in
a {\em generic symplectic} basis {\bf z}.  Under a symplectic
change of basis ${\bf z} \to {\bf z}' = {\bf z}S,\;S \in
Sp(2n,{\cal R})$, the matrix $M({\bf z})$ undergoes the
following transformation:
\begin{equation}
M({\bf z})\to M({\bf z}^\prime) =  S^TM({\bf z})S\,,\,\,\, S \in
Sp(2n,R)\,.
\end{equation}
Since $S^T \beta S = \beta$ implies $\beta S^{T} = S^{-1}
\beta$, we have
\begin{equation}
S:~~ \beta M({\bf z}) \to \beta M({\bf z}^\prime) =  S^{-1}\beta
M({\bf z}) S.
\end{equation}
That is, under a symplectic change of basis $ M({\bf z})$
undergoes congruence, but $\beta M({\bf z})$ undergoes
conjugation. Hence $\mbox{tr}(\beta M({\bf z}))^{2l},\,\,
l=1,2,\cdots,n$ are  invariant (Note that  ${\rm tr}(\beta
M({\bf z}))^{2l+1}=0$ in view of $\beta ^T = -\beta,
\;M({\bf z})^T = M({\bf z})$).

Since $i\beta$ is an {\em involution} we can use it to separate
$M({\bf z})$ into even and odd parts :
\begin{eqnarray}
M({\bf z})&=&M({\bf z})^{\rm even} + M({\bf z})^{\rm
odd}\,,\nonumber\\ M({\bf z})^{\rm even} &=& {1\over2}(M({\bf
z})+\beta M({\bf z})\beta^T)\,,\nonumber\\ M({\bf z})^{\rm odd}
&=& {1\over2} (M({\bf z})-\beta M({\bf z}) \beta^T)\,.
\end{eqnarray}
The even and odd parts of $M({\bf z})$ satisfy the symmetry
properties
\begin{equation}
\beta  M({\bf z})^{\rm even}\beta ^T=M({\bf z})^{\rm
even}\,,\,\,\,\beta M({\bf z})^{\rm odd}\beta^T = -M({\bf
z})^{\rm odd}\,.
\end{equation}
Further, $M({\bf z})^{\rm odd}$ and $M({\bf z})^{\rm even}$ are
trace orthogonal: $\mbox{tr}\left(M({\bf z})^{\rm odd}M({\bf
z})^{\rm even}\right) = 0$.

The structure of the even and odd parts of $M({\bf z})$ may be
appreciated by writing  $M({\bf z})$ in the block form
\begin{equation} M({\bf z})=\left(\begin{array}{cc} A & C\\ C^T
& B\,\end{array}\right),\,\,A^T=A\,,\,\,B^T=B\,.
\end{equation}
We have
\begin{eqnarray}
M({\bf z})^{\rm even} &=& \left(\begin{array}{cc} {1\over2}(A+B)
& {1\over2}(C-C^T)\\ &  \\ -{1\over2}(C-C^T) &
{1\over2}(A+B)\end{array}\right),\nonumber\\ \nonumber\\ M({\bf
z})^{\rm odd} &=& \left(\begin{array}{cc} {1\over2}(A-B) &
{1\over2}(C+C^T)\\ &  \\ {1\over2}(C+C^T) &
{1\over2}(B-A)\end{array}\right).
\end{eqnarray}

Now consider the invariant $-{\rm tr} (\beta M({\bf z})\beta
M({\bf z}))={\rm tr}(\beta ^T M({\bf z})\beta M({\bf z}))$. We
have
\begin{equation}
{\rm tr}(\beta^T M({\bf z})\beta M({\bf z})) = {\rm tr}(M({\bf
z})^{\rm even})^2 - {\rm tr}(M({\bf z})^{\rm odd})^2\,,
\end{equation}
which can also be written as
\begin{equation}
{\rm tr}(\beta^TM({\bf z})\beta M({\bf z})) = {\rm tr}(M({\bf
z})^2) - 2{\rm tr}(M({\bf z})^{odd})^2\,.
\end{equation}
The structural similarity of this invariant to that in the
pseudo-orthogonal case should be appreciated.

Now, by an argument similar to the pseudo-orthogonal case one
finds that, owing to the noncompactness of $Sp(2n,{\cal R})$,
the function $\mbox{tr}(M({\bf z})^2)$ and hence the
Schweinler-Wigner quartic form $\displaystyle{\sum_{k=1}^{2n}}
(M({\bf z})_{kk})^2$  is unbounded  if ${\bf z}$  is allowed to
run over the entire manifold of all symplectic bases. For
instance,  in the lowest dimensional case $n=1$ with $M$ chosen
to be
\begin{equation}
M= \left(\,\begin{array}{cc}a & u\\d & b\,\end{array}\right)
,~~a, b > 0,~~ab-ud>0,
\end{equation} 
under congruence by the $Sp(2,{\cal,R})$ matrix
\begin{equation}
S= \left(\,\begin{array}{cc}
\mu & 0\\
0 & 1/\mu\,\end{array}\right),
\end{equation}
the value of $\displaystyle{\sum_k} (M_{kk})^2$ changes from
$a^2 + b^2$ to $ \mu^2 a^2 +(1/\mu^2) b^2 $ which, by an
appropriate choice of $\mu$, can be made as large as one wishes.

However, it follows from (41) that over the submanifold of
symplectic bases which leave ${\rm tr}(M({\bf z})^{\rm odd})^2$
fixed, the function ${\rm tr}(M({\bf z})^2)$ remains invariant
and so the quartic form $\sum (M({\bf z})_{kk})^2$ is bounded
within this restricted class of symplectic bases and hence can
be maximised.  In particular the nonnegative ${\rm tr}(M({\bf
z})^{\rm odd})^2$ can be chosen to take its minimum value.
Williamson theorem  implies that there are symplectic bases
which realize the absolute mimumum ${\rm tr}(M({\bf  z})^{\rm
odd})^2 = 0$.

We can now formulate the {\em analogue of the Scweinler-Wigner
extremum principle for symplectic bases} in the following way:
Take the subfamily of symplectic bases in which ${\rm tr}(M({\bf
z})^{\rm odd})^2$ and hence ${\rm tr}(M({\bf z})^2)$is minimum.
[This minimum of $\mbox{tr}(M({\bf z})^2)$ equals the invariant
${\rm tr}(\beta^TM({\bf z})\beta M({\bf z}))$]. Then maximise
the Schweinler-Wigner quartic form $ m({\bf z}) =
\displaystyle{\sum_k} (M({\bf z})_{kk})^2$ within this
submanifold of symplectic bases. This will lead, not just to a
basis in which $M({\bf z})$ is diagonal, but to one where
$M({\bf z})$ has the Williamson canonical form $M({\bf z})={\rm
diag}(\kappa_1,\cdots, \kappa_n;
\kappa_1,\cdots, \kappa_n)$. We have thus established the following 
generalization of the Schweinler-Wigner extremum principle to
the symplectic case.

\noindent
{\em Theorem} 7: The distinquished symplectic basis which
extremizes the ``Schweinler-Wigner'' quartic form $m({\bf z})$
over the submanifold of symplectic bases which minimize the
quartic form $\mbox{tr}(M({\bf z})^2)$ is the same as the
symplectic basis in which the positive definite matrix $M({\bf
z})$ assumes the Williamson canonical diagonal.

Note that once $M({\bf z})^{\rm odd}=0$ is reached, as implied
by ${\rm tr}(M({\bf z})^{\rm odd})^2=0$, $M({\bf z})$ has the
special even form
\begin{equation}
{\left(\begin{array}{cc} A & C\\ -C &
A\end{array}\right)},~~A^T=A,\;C^T = -C,
\end{equation}
so that $A+iC$ is hermitian. The subgroup of symplectic
transformations which do not mix $M({\bf z})^{\rm even}$ with
$M({\bf z})^{\rm odd}$, and hence maintain the property $M({\bf
z})^{\rm odd}=0$ have the special form
\begin{equation}
S = \displaystyle{\left(\begin{array}{cc} X & Y\\ -Y &
X\end{array}\right)},~~~ X+iY \in ~U(n).
\end{equation}
This subgroup, isomorphic to the unitary group $U(n)$, is the
maximal compact subgroup\cite{dutta} of $Sp(2n,{\cal R})$. Thus,
diagonalizing $M({\bf z})$ using symplectic change of basis,
after it has reached the even form, is the same as diagonalizing
an $n$-dimensional hermitian matrix using  unitary
transformations.

\section{\bf Concluding Remarks}

To conclude, we have shown that an  $N\times N$ real symmetric
[complex hermitian] positive definite matrix is congruent to a
diagonal form modulo a pseudo-orthogonal [pseudo-unitary] matrix
belonging to $SO(m,n)$ [$SU(m,n)$], for any choice of partition
$N=m+n$. The method of proof of this result is adapted to
provide a simple proof of Williamson's theorem. An important
consequence of these theorems is that while a real-symmetric
[complex-hermitian] positive definite matrix has a unique
diagnal form under conjugation, it has several different
canonical diagnal forms under congruence.  The theorems
developed here are used to formulate an extremum principle a
l\'a Schweinler and Wigner for constructing
pseudo-orthonormal[pseudo-unitary] and symplectic bases from a
given set of linearly independent vectors. Conversely, the
extremum principle thus formulated can be used for finding the
congruence transformation which brings about the desired
diagonalization.

It is interesting that pseudo-orthonormal basis and symplectic
basis could be constructed by extremizing {\em precisely the
same Schweinler-Wigner quartic form} $ m({\bf z}) =
\displaystyle{\sum_k} (M({\bf z})_{kk})^2$ that was originally
used to construct orthonormal basis in the unitary case.
However, it must be borne in mind that the similarity in the
structure of the quartic form to be extremized in the three
cases considered is only at a formal level. In reality, the
three quartic forms are very different objects, for they are
functions over topologically very different manifolds: ${\bf z}$
runs over the group manifold $U(N)$ of orthogonal frames in the
original Schweinler-Wigner case, the group manifold $SO(m,n)$ of
pseudo-orthogonal frames in the Lorentz case, and over the group
manifold $Sp(2n,{\cal R})$ in the symplectic case. This has the
consequence that, unlike the orthogonal case, this quartic form
is unbounded in the noncompact $SO(m,n) [SU(m,n)]$ and $Sp(2n,
{\cal R}) [Sp(2n, {\cal C})] $ cases. Insight into the structure
of these groups was used to achieve constrained extremization
within a natural maximal compact submanifold.  


\begin{references}
\bibitem{[6]} 
Just as diagonal form is the canonical form for real symmetric
matrices uner rotation, $i\sigma_2\bigotimes K$, with $K$
diagonal, is the canonical form for a real antisymmetric matrix
under rotation. Further $K$ can be chosen to be non-negative, in
general, and positive definite when the antisymmetric matrix is
nonsingular.
\end{references}
\end{document}